\documentclass[prc,aps,preprint,superscriptaddress,showpacs]{revtex4-1}
\usepackage{amssymb}
\usepackage[dvips]{graphicx}
\usepackage[english]{babel}
\usepackage{indentfirst}
\usepackage{amsxtra}
\usepackage{amsmath}
\usepackage{supertabular}
\usepackage{multirow}
\usepackage [mathcal]{eucal}
\begin{document}
\title {\bf Probing the pairing interaction through two-neutron   
transfer reactions}

\author{E. Pllumbi}
\affiliation{Dipartimento di Fisica, University of Pisa, Largo Bruno Pontecorvo 3, 56127 Pisa, Italy}

\author{M. Grasso}
\affiliation{Institut de Physique Nucl\'eaire, IN2P3-CNRS, Universit\'e Paris-Sud, 
F-91406 Orsay Cedex, France}

\author{D. Beaumel}
\affiliation{Institut de Physique Nucl\'eaire, IN2P3-CNRS, Universit\'e Paris-Sud, 
F-91406 Orsay Cedex, France}

\author{E. Khan}
\affiliation{Institut de Physique Nucl\'eaire, IN2P3-CNRS, Universit\'e Paris-Sud, 
F-91406 Orsay Cedex, France}

\author{J. Margueron}
\affiliation{Institut de Physique Nucl\'eaire, IN2P3-CNRS, Universit\'e Paris-Sud, 
F-91406 Orsay Cedex, France}

\author{J. van de Wiele}
\affiliation{Institut de Physique Nucl\'eaire, IN2P3-CNRS, Universit\'e Paris-Sud, 
F-91406 Orsay Cedex, France}

\begin{abstract}

Cross sections for ($p,t$) two-neutron transfer reactions are calculated 
in the one-step zero-range 
distorted-wave Born approximation for the tin isotopes 
$^{124}$Sn and $^{136}$Sn and for incident proton energies 
from 15 to 35 MeV. Microscopic quasiparticle 
random-phase approximation 
form factors are provided
for the reaction calculation and phenomenological optical potentials 
are used in both the entrance 
and the exit channels. Three different surface/volume mixings of   
a zero-range density-dependent 
pairing interaction are employed in the microscopic calculations and the sensitivity of the 
cross sections to the
different mixings is analyzed. Since absolute cross sections cannot be obtained 
within our model, we compare the positions of the diffraction 
minima and the 
shapes of the angular distributions. No differences are found in the position of the diffraction 
minima for the reaction   
$^{124}$Sn($p,t$)$^{122}$Sn. On the other side,    
the angular distributions obtained 
for 
the 
reaction $^{136}$Sn($p,t$)$^{134}$Sn
with surface and mixed 
interactions differ at large angles for some values of the incident proton energy. 
For this reaction,  
we compare the ratios of the cross sections associated 
to the ground state and the first excited state transitions. Differences among the three 
different theoretical predictions are found and they are 
more important at the incident proton energy of 15 MeV.  
As a conclusion, 
we indicate ($p,t$) two-neutron transfer reactions with 
very neutron-rich Sn isotopes and at proton energies around 15 MeV as 
good experimental cases where the surface/volume mixing of the 
pairing interaction may be probed. 

\end{abstract} 

\vskip 0.5cm  \maketitle

%-----------------------------------------------------------------------
\section{Introduction}

Pairing correlations play an important role in determining the 
properties  of a large number of open-shell superfluid nuclei. 
Many efforts have been devoted in the last years   
to investigate and better clarify the deep nature 
of the pairing interaction in several nuclear systems, from finite atomic nuclei to 
compact neutron stars, 
which may be regarded as the most exotic 
nuclear systems in nature. A detailed review of methods, analyses and results 
concerning    
the pairing interaction and the study of 
pairing correlations has been published in 
2003 by Dean and Hjorth-Jensen    
\cite{morten}. We mention in what follows 
some examples of more recent works which are devoted to   
the study of nuclear superfluidity: (i) 
the localization of Cooper pairs in the 
nuclear medium has been analyzed within different models 
and in different contexts  
\cite{pillet,mat05,hag07,mat06,hag207}; 
(ii) polarization effects, i.e., 
the impact on pairing correlations of the coupling to collective phonons is 
discussed in the community (see, for example, Refs. \cite{barranco,potel1}); 
(iii)  
a bridge between finite nuclei and infinite matter has been established and 
the pairing gaps in 
nuclei have been calculated by fitting the interaction on 
symmetric and neutron  
matter \cite{mar08}; (iv) $T=1$ and $T=0$ 
(neutron-proton pairing in $N \sim Z$ nuclei) are being investigated and,  
in this respect, an interesting study of partial-wave 
contributions to pairing has been recently presented by Baroni {\it et al.} 
\cite{baroni}. 
 
The derivation of the pairing interaction on a fully microscopic basis 
is being 
performed nowadays and pairing gaps have actually been  
obtained from low-momentum 
interactions $V_{low-k}$ \cite{lesi}. 
However, in most of the available 
mean-field-based models, a more phenomenological attitude is still usually 
adopted also because the agreeement between the 
theoretical $V_{low-k}$-gaps and the experimental values does not     
 improve significantly with respect to  
what obtained with phenomenological interactions. 
It is worth mentioning that 
in phenomenological interactions 
not only the bare interaction but also higher-order 
terms are taken into account in an effective way. 
While in the Gogny case almost the same interaction is employed in both 
particle-hole and particle-particle channels,  
in Skyrme-mean-field-based models the pairing interaction which is used 
in the particle-particle channel is usually 
different from the interaction used in the mean-field 
channel \cite{garri}. This prevents the problems related to double counting 
in the particle-particle channel. 
One of the current choices for the pairing interaction is a zero-range interaction   
depending on the isoscalar density. Extensions of this form to include also 
a dependence on the isovector density have been recently proposed \cite{mar07,yama}. 
The parameters appearing in the expression of the pairing interaction  
are fitted on nuclear properties 
following different criteria. One of these criteria consists in using the 
experimental odd-even mass staggering as a constraint and this 
choice has been extensively analyzed 
\cite{doba,gori,doba2,duguet}.

The role of the pairing vibration modes 
(associated to pair-transfer reactions) 
in providing a helpful insight into  
superfluidity in nuclei has been 
discussed since several years \cite{bro,vittu}. 
For recent reviews on the main advances achieved in multinucleon 
transfer reactions at energies close to the Coulomb barrier 
see Refs. \cite{corradi1,corradi2}. 
The excitation modes related to the 
transfer of nucleonic pairs  
in superfluid nuclei are actually expected to be strongly sensitive 
to the specific features of the pairing interaction \cite{vittu}.
It has been recently suggested that 
the observables related to pairing vibrations could be 
considered as useful 
additional constraints in the fitting procedures of 
phenomenological interactions \cite{kh09,mat09}. In particular, in the 
currently used
 zero-range density-dependent interaction, the surface/volume mixing 
of the interaction can be tuned by modifying the parameter $x$ in the 
expression
\begin{eqnarray}
V(\vec{r}_1, \vec{r}_2)= \delta (\vec{r}_1 - \vec{r}_2)
V_0 \left[ 1+x \left( \frac{\rho(\vec{R})} {\rho_0} \right) 
^{\alpha} \right]~,
\label{eq1}
\end{eqnarray}
where $\vec{R}=(\vec{r}_1 + \vec{r}_2)/2$; $x=0$ and $x=1$ 
represent the extreme cases of a pure volume and a pure 
surface interaction, 
respectively.  Microscopic quasiparticle 
random-phase approximation (QRPA) calculations for the 0$^+$ \cite{kh09} and 
the 2$^+$ \cite{mat09} pair-transfer modes have been performed and it has been 
shown that the transition densities actually 
depend on the different choices of the 
pairing interaction in terms of surface/volume mixing. 
In particular, differences have been found between the two cases 
of a pure surface interaction and a mixed interaction. 

To finally prove that 
the observables associated to pairing vibrations can indeed guide us 
toward a 
deeper comprehension of this specific aspect of the pairing 
interaction,  cross sections have to be evaluated. 
The analysis is pursued here in this direction. 
The idea is to perform a calculation of cross sections where the form factor 
of the transition is evaluated microscopically. 
The self-consistent microscopic QRPA 
results are used as structure 
inputs for the reaction calculation. 
We consider two-neutron tranfer ($p,t$) reactions where the 
Hartree-Fock-Bogoliubov (HFB) + QRPA approach is used to describe the 
microscopic nuclear structure and the one-step 
distorted-wave Born approximation 
(DWBA) is employed for the reaction dynamics. The cross sections in the DWBA 
are calculated using the Distorted Waves University of Colorado Kunz 4 
(DWUCK4) code \cite{dwuck}. While the evaluation of the form 
factors is based on self-consistent and 
completely 
microscopic structure calculations, some 
limitations of the present reaction calculations have to be mentioned.  
A first limitation is the inability of such calculations to reproduce 
measured absolute cross sections \cite{franey}: 
we thus compare in this work only 
angular profiles and in particular 
the location of the 
diffraction minima. Furthermore, 
it has to be noticed that in 
one-step DWBA calculations inelastic excitations in the reaction channels and two-step 
processes corresponding to sequential particle transfers are missing 
\cite{franey}. 
The relative importance of these processes is still quite an open question. 
Owing to these limitations, we consider the results presented in this work as 
a first qualitative indication and not a precise prediction 
about the sensitivity of the 
cross sections to the surface/volume character of the pairing 
interaction. 

The analysis presented in this work is done 
for the tin isotopes $^{124}$Sn, which is stable, and $^{136}$Sn, which is 
unstable and very neutron-rich. 

The article is organized as follows. In Sec. II the microscopic QRPA form 
factors are shown. 
In Sec. III  
the cross sections are analyzed 
and a comparison with the experimental data is presented for the case 
$^{124}$Sn($p,t$)$^{122}$Sn. 
Conclusions and  
perspectives are summarized in Sec. IV.

\section{Microscopic form factors}
\label{input}

For the excitation modes associated to ($p,t$) reactions, 
the transition densities are calculated between the ground state of the 
nucleus with $A$ nucleons and a state (the ground state or an excited state) 
of the nucleus with $A-2$ nucleons. 
The form factor is obtained by folding the transition density with the 
interaction between the transferred pair and the residual nucleus. In our 
model, this interaction is 
actually chosen of zero-range and this means that the transition densities 
directly provide the form factors. It is clear that with a zero-range 
interaction the shapes of the angular distributions can be well defined but 
the absolute values of the cross sections cannot be evaluated 
\cite{kh04}. 

The structure calculations to derive the transition 
densities are performed in this work within the HFB + QRPA framework 
\cite{kh04,kh02}. In this model, the modes associated to 
the transfer of pairs are obtained by considering the particle-particle 
(hole-hole) components of the QRPA Green's function for the transition 
$A \rightarrow A+2$ ($A \rightarrow A-2$) \cite{kh04}.
The strength function describing an excitation in the 
particle-hole channel is given by the well-known expression 
\begin{eqnarray}
S(\omega)= - \frac{1}{\pi} Im \int F^{11*} (r) 
G^{11}(r,r';\omega) F^{11}(r') dr dr'~,
\label{eq2}
\end{eqnarray}
where '1' denotes the particle-hole subspace. $G^{11}$ and $F^{11}$ are 
thus the 
particle-hole components of the QRPA Green's function and of the 
excitation operator, respectively. 

Similarly, the strength function for 
the transition  ($A \rightarrow A+2$) is written as 
\begin{eqnarray}
S(\omega)= - \frac{1}{\pi} Im \int F^{12*} (r) 
G^{22}(r,r';\omega) F^{12}(r') dr dr'~,
\label{eq3}
\end{eqnarray}
where '2' denotes the particle-particle subspace.   
 The strength function describing  
the transition  ($A \rightarrow A-2$) is
\begin{eqnarray}
S(\omega)= - \frac{1}{\pi} Im \int F^{13*} (r) 
G^{33}(r,r';\omega) F^{13}(r') dr dr'~,
\label{eq4}
\end{eqnarray}
where '3' represents the hole-hole subspace. 
In the present QRPA calculations, the 
strength distributions have been 
evaluated using Eq. (\ref{eq4}). 

The first peak of the response functions given by Eqs. 
(\ref{eq3}) and (\ref{eq4}) describes the transition from the 
ground state of the $A$ nucleus to the ground state of the 
$A \pm 2$ nucleus.

The pair transition density is given by the following expression: 
\begin{eqnarray}
\delta \kappa^{\nu}(r \sigma)= \langle 0 | c(r \tilde{\sigma}) 
c(r \sigma) |\nu \rangle ~,
\label{eq5}
\end{eqnarray}
where $\nu$ is the state under consideration (either the ground 
state or an excited state of the final nucleus).  

As already done 
in Ref. \cite{kh09}, we use three different values of the parameter $x$ 
in Eq. (1), $x=$ 0.35, 0.65 and 1. The first two cases are associated to 
a mixed pairing interaction while $x=1$ corresponds to a pure surface 
interaction.  The other parameters of the 
pairing interaction are adjusted in the same way as in Ref. \cite{kh09} 
(the fit is done on the two-neutron separation energies of Sn isotopes). 
The only difference between the present structure calculations and those 
shown in Ref. \cite{kh09} is that the transition  
$A$ $\rightarrow$ $A-2$ transition is considered here 
whereas  the transition $A$ $\rightarrow$ $A+2$ 
has been explored  in Ref. \cite{kh09}.  

In this work, we are interested in the transitions to the 
first 0$^+_1$ state, which is the 
ground state ($gs$) of the $A-2$ nucleus, 
and 
to the first 0$^+$ excited state (0$^+_2$) of the 
final $A-2$ nucleus. We thus consider the first two peaks of the 
QRPA response functions and calculate the corresponding transition 
densities. 
The transition densities for the $^{124}$Sn($p,t$)$^{122}$Sn and  
$^{136}$Sn($p,t$)$^{134}$Sn reactions 
are shown in Figs. 1 and 2: Left panels refer to the $gs$ 
transition and right panels refer to the 0$^+_2$ transition. 
%({\it figures 4.1 and 4.2 extracted from Else thesis, pages 53 and 54. 
%Good quality figures to be done}). 
In general, one observes that 
the two form factors associated to the two mixed interactions are 
quite similar one to the other and different from 
the form factor  corresponding to 
$x=1$, especially in the right panels (0$^+_2$ 
transitions). Similar results have been already found in Ref. \cite{kh09}.

\section{Cross sections with different surface/volume mixings}
\label{cross}
 
Optical potentials have to be provided together with the form factors 
for the reaction calculations. Phenomenological optical potentials 
are used here. The optical potential for the entrance channel 
(interaction between the proton and the nucleus) is constructed with the 
parameters of Ref. \cite{becchetti} which have been fitted on 
elastic scattering of protons by 
nuclei with $A > 40$ and with a proton laboratory energy $E_p < 50$  
MeV. For the exit channel, the optical potential parameters have been 
fitted on the elastic scattering of the triton \cite{li}. 

In the DWBA calculations performed with the DWUCK4 code, a neutron pair 
of zero angular momentum is transferred from one nucleus to the other. 
The $gs$ $\rightarrow$ $gs$ 
 and the $gs$  $\rightarrow$ 0$^+_2$ 
transitions are calculated for different incident proton energies $E_p$ 
and with the three different pairing interactions.   
Incident proton energies range from 15 to 35 MeV. 

Some experimental data for the reaction $^{124}$Sn($p,t$)$^{122}$Sn are available 
\cite{bassani,matoba}. We have compared the theoretical angular distributions 
with the 
experimental points (Fig. 3) for the $gs$ transition at $E_p =$ 20 MeV (left) 
and for the 0$^+_2$  transition at $E_p = $ 35 MeV (right). 
Since the  
calculated cross sections are not absolute, we have normalized them at the 
experimental amplitudes. The three curves are 
normalized to reproduce the highest 
experimental maximum in the two panels. 
Owing to this normalization procedure, we cannot 
compare the absolute values but only the 
angular profiles, i.e., the position of the diffraction minima. 
It turns out that for the 
three used pairing interactions the minima are located in all cases at the same 
values of the angle $\Theta_{CM}$ (in agreeement with the experimental data). 
One cannot thus deduce in this case any helpful hint about the surface/volume 
mixing of the pairing interaction.  

When drip lines are approached, neutron skins become thicker:  
surface effects and low-density pairing features are thus 
expected to be typically more important. Owing to this, 
we have considered also the 
pair-transfer reaction $^{136}$Sn($p,t$)$^{134}$Sn where a more neutron-rich 
nucleus is involved. 
With the next-generation facilities, it is expected that beams of very 
neutron-rich tin isotopes such as $^{134,136}$Sn will be produced with 
sufficiently high intensity for performing two-nucleon tranfer 
experiments. 
We have performed the same kind of calculations as for the case 
$^{124}$Sn($p,t$)$^{122}$Sn 
($E_p $ ranging from 15 to 35 MeV). 
In this case, we have checked the sensitivity of the calculations 
with respect to the choice of the optical potential in the 
entrance channel. We have found similar results using the optical 
potential of Ref. \cite{varner}. 

For some values of the proton incident energy, the location of the 
diffraction minima is not the same when different pairing 
interactions are used. 
We show in Fig. 4 only the relevant cases, 
i.e., those corresponding to the incident energies for which 
some discrepancies have been found in the angular profiles 
associated to different pairing 
interactions. The two cases of mixed interactions 
($x = $ 0.35 and $x=$ 0.65) do not actually 
significantly differ one from the other while discrepancies are found between the pure surface 
case and the mixed cases. As an illustration, we show for the mixed interaction only the case 
$x=$ 0.35. We compare in Fig. 4 the results obtained for $x=$ 1 and $x=$ 0.35.  
In the left panel, the $gs$ transition is described for the two values of incident energy 
$E_p = $ 30 and 35 MeV. 
For $E_p=$ 30 MeV, the curves to compare are the solid black ($x=1$) and the 
dashed green ($x=0.35$). For $E_p=$ 35 MeV, the curves to compare  
are the solid red ($x=1$) and the 
dashed blue ($x=0.35$). 
In the right panel, the 0$^+_2$ transition is described for  
$E_p = $ 15 MeV. One observes in both panels that the profiles of the cross sections 
corresponding to $x=$ 1 and $x=$ 0.35 differ at large angles 
($\Theta_{CM} >$ 70 degrees). We are aware that measurements are more difficult at 
large angles because the corresponding cross sections are 
very low. 
Nevertheless, this result indicates that very neutron-rich Sn isotopes may be 
interesting cases to analyze. 
On the basis of this first indication, 
we continue our investigation for $^{136}$Sn  
and we show in Fig. 5 the ratios of the cross sections associated to the 
$gs$ and to the 0$^+_2$ 
transitions at different proton energies. 
Even if absolute cross sections cannot be calculated within the present reaction model, the ratios of 
the cross sections related to the $gs$ and the 0$^+_2$ 
transitions are meaningful quantities to analyze. 
These ratios are proportional to the ratios of the transition 
probabilities associated to the two transitions and that the proportionality factor is the same 
independently of the  
pairing interaction. The comparison of the ratios obtained with different pairing interactions can thus provide 
interesting predictions about the sensibility 
of the cross sections 
to the choice of the pairing interaction. 
It can be seen that differences exist 
among the three sets of results and they are more important at the lowest energy of 
15 MeV, that represents the case where the reaction takes place mostly 
in the surface region of the nucleus. Hence, we suggest 
very neutron-rich Sn isotopes 
and proton energies around 15 MeV 
as favorable cases for future ($p,t$) or 
($t,p$)
pair-transfer experiments  
that can provide a deeper insight into the surface/volume character 
of the pairing interaction. 
Performing ($t,p$) reaction masurements in inverse kinematics is quite more 
challenging than ($p,t$), but such reactions would allow one to populate different
states of Sn isotopes that may also represent very favourable cases for 
pairing studies.  

\section{Summary and Perspectives}
We have evaluated ($p,t$) two-neutron transfer-reaction cross sections in the 
zero-range DWBA approximation for the two cases $^{124}$Sn($p,t$)$^{122}$Sn and 
$^{136}$Sn($p,t$)$^{134}$Sn. 
The limitations of these reaction calculations have been 
discussed. 
Microscopic form factors have been provided for the 
reaction calculation where the transition densities are calculated within the 
HFB+QRPA approach. A phenomenological zero-range density-dependent pairing 
interaction is used in the particle-particle channel of this model and three 
different surface/volume mixings have been considered for the pairing 
interaction: a pure surface interaction and two mixed surface/volume 
interactions. The sensitivity of the cross sections to these different choices 
has been investigated for the two reactions. Our reaction calculation 
does not provide absolute cross sections. We thus compare only the 
angular profiles and in particular   
the location of the diffraction minima. We consider several values of the incident 
proton energy, from 15 to 35 MeV. 

For $^{124}$Sn($p,t$)$^{122}$Sn, negligible discrepancies are found in the 
location of the minima and in the shape of the angular distributions among the 
different pairing cases for all the considered 
incident proton energies. For this reaction, 
experimental data exist at $E_p=$ 20 and 35 MeV and the comparison with the 
experimental points is satisfactory in all cases, independently of the  
the pairing interaction. 

A more interesting case seems to be the reaction 
$^{136}$Sn($p,t$)$^{134}$Sn where a very neutron-rich nucleus is 
involved. For 
some values of the energy of the proton, 
discrepancies at large angles ($\Theta_{CM} > $ 
70 degrees) are found in the position of the minima between the 
pure-surface case and the 
mixed-interaction case (the two mixed interactions lead 
to very similar results). 
This is a first qualitative indication that suggests that this case 
can be interesting to be explored experimentally. 
We have compared the ratios of the cross sections associated to the 
$gs$ and to the 0$^+_2$ transitions at different proton energies. 
Sizeable 
differences among the three theoretical predictions are found especially
at the lowest proton energy of 15 MeV. 
New-generation accelerators should allow soon 
to produce $^{134,136}$Sn beams with sufficiently high 
intensity.  
The conclusion of this work is  
that pair-transfer reactions for 
a very neutron-rich Sn isotope 
and at proton energies around 15 MeV (reactions in the surface 
region of the nucleus) 
may be good experimental  
cases where the surface/volume nature of the pairing interaction can be 
elucidated. This is a first indication.  
More precise reaction calculations should certainly be performed to 
get absolute cross sections; more accurate predictions 
could be obtained by 
taking into account more complex processes 
like two-step excitations 
which are so far neglected in the present calculations 
whereas they are included in more sophisticated reaction codes 
\cite{potel}.

%\begin{acknowledgments}

\begin{acknowledgments}
The authors thank O. Sorlin for fruitful 
 discussions.
\end{acknowledgments}
%

%\end{acknowledgments}
%
\newpage
%\bibliographystyle{revtex.bst} 
%\bibliography {biblioTensor}
%
%
\providecommand{\noopsort}[1]{}\providecommand{\singleletter}[1]{#1}%

\clearpage
\newpage
%
%
%
%-----------------------------------

%
\begin{figure}[htb]
\begin{center}
\parbox[c]{16cm}{\includegraphics[width=14cm]{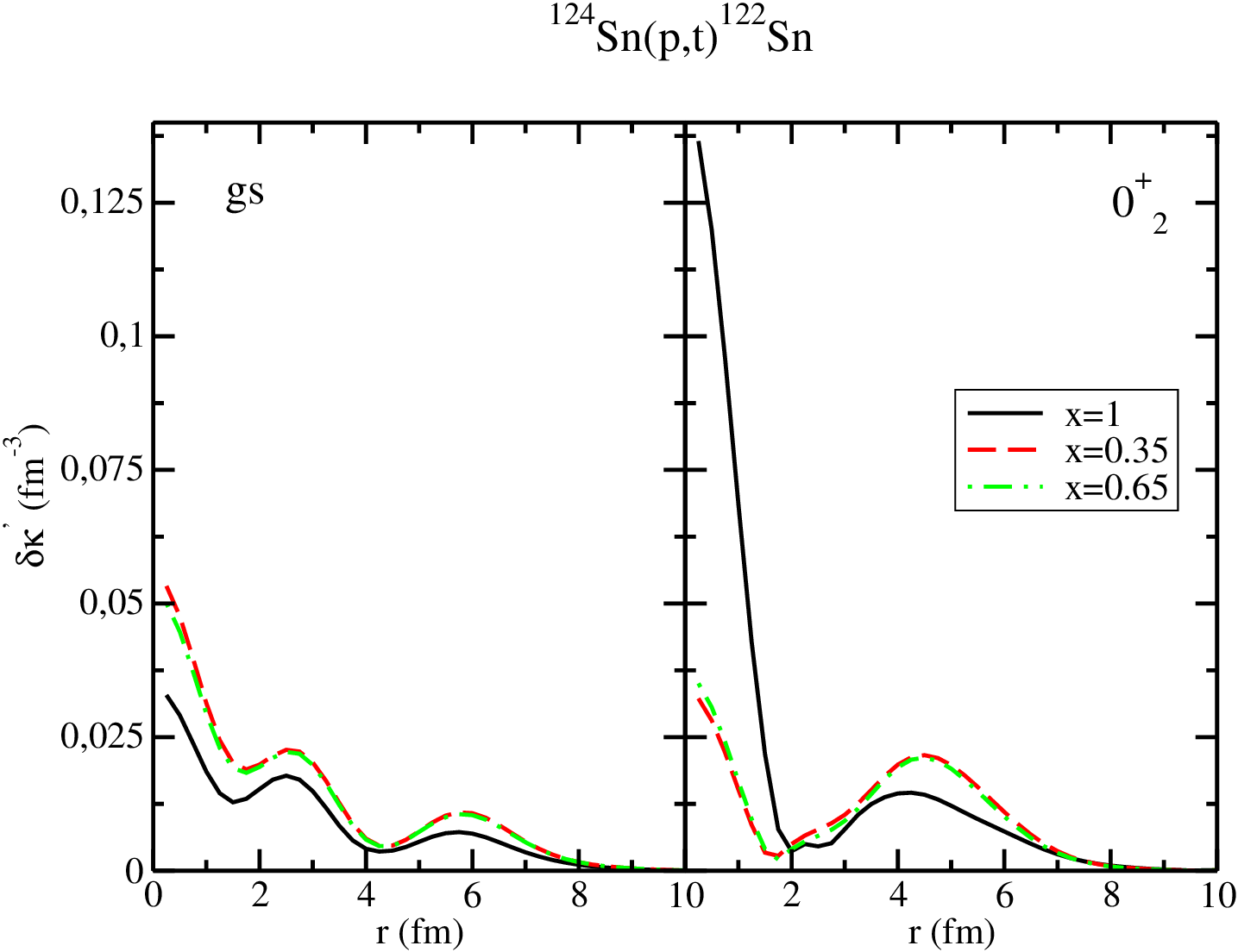}}
\end{center}
\caption{\small (Color online) Form factors of the reaction 
$^{124}$Sn($p,t$)$^{122}$Sn for the transition 
to the ground state (left) and 
to the first 0$^+$ excited state (right).}
\label{fig1}
\end{figure}
\begin{figure}[htb]
\begin{center}
\parbox[c]{16cm}{\includegraphics[width=14cm]{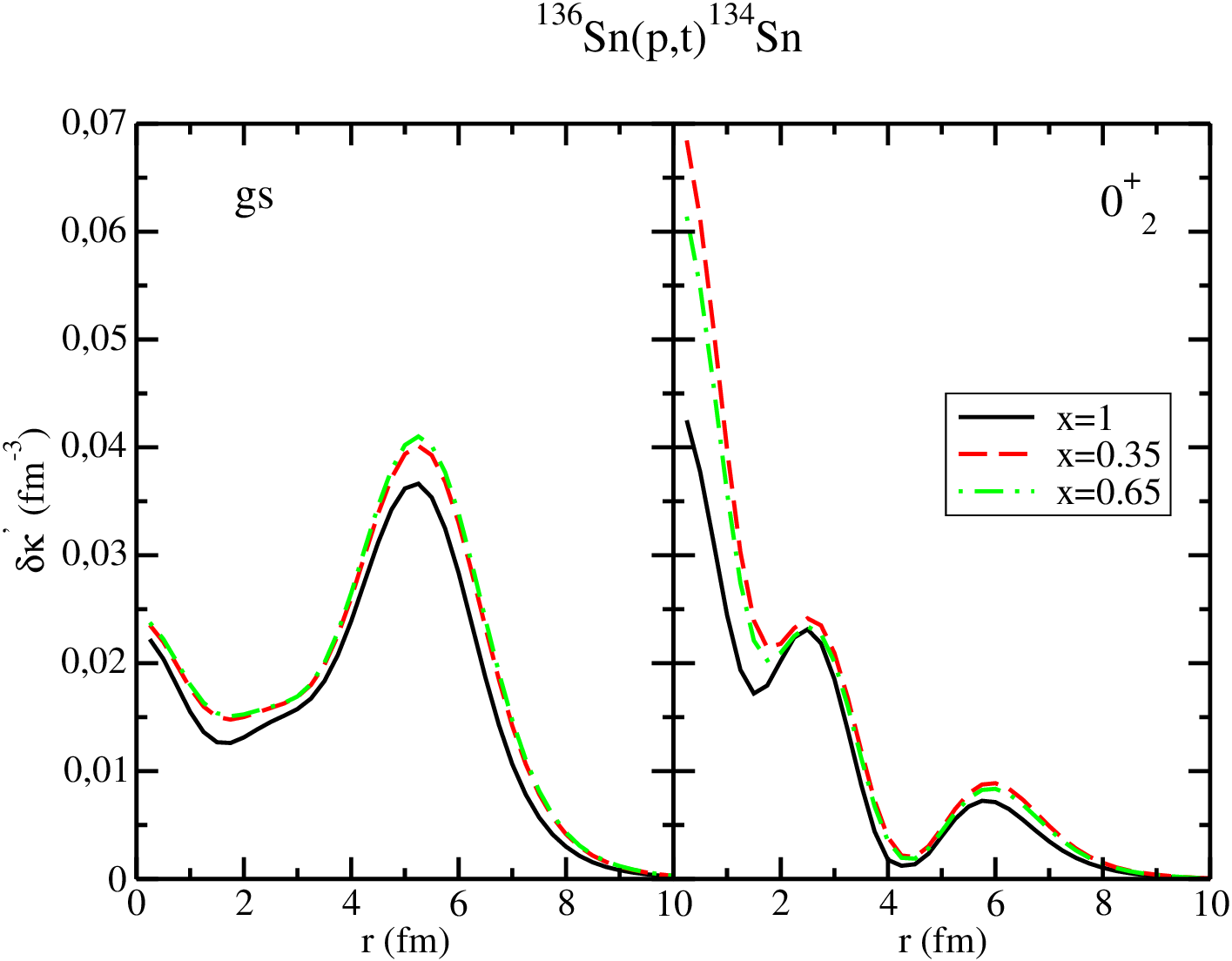}}
\end{center}
\caption{\small (Color online) Form factors of the reaction 
$^{136}$Sn($p,t$)$^{134}$Sn for the transition 
to the ground state (left) and 
to the first 0$^+$ excited state (right).}
\label{fig2}
\end{figure}

%\begin{figure}[htb]
%\begin{center}
%\parbox[c]{16cm}{\includegraphics[width=14cm]{fig3-else.eps}}
%\end{center}
%\caption{\small (Color online) QRPA form factors 
%(black solid lines} and corrected ones 
%(red dashed lines) for the $^{124}$Sn($p,t$)$^{122}$Sn 
%reaction 
%with the three different choices 
%of the pairing interaction.}
%\label{fig3}
%\end{figure}

\begin{figure}[htb]
\begin{center}
\parbox[c]{16cm}{\includegraphics[width=14cm]{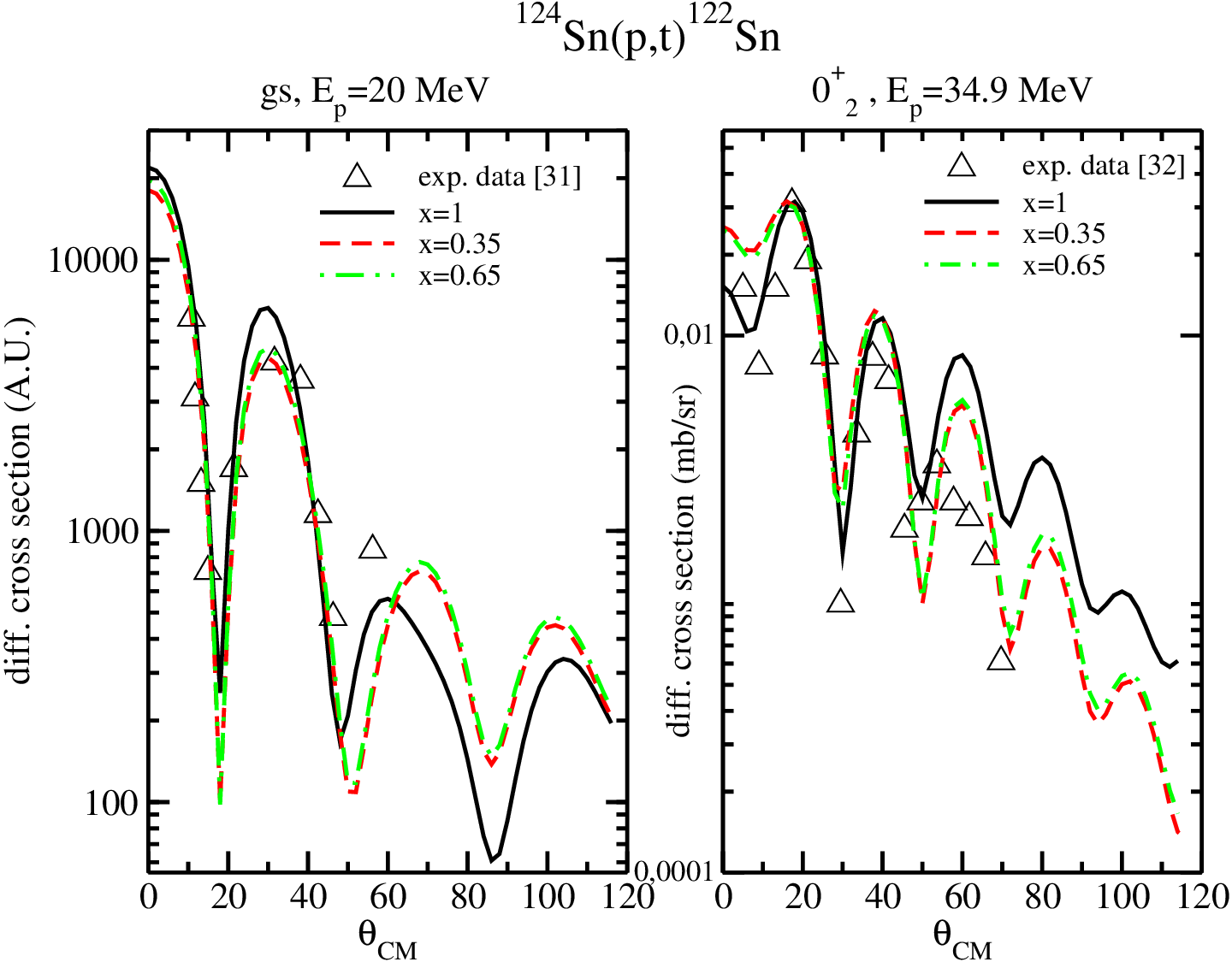}}
\end{center}
\caption{\small (Color online) Left: Comparison between the calculated and 
the experimental cross sections for the transition to the ground state of the final
 nucleus. The reaction is 
$^{124}$Sn($p,t$)$^{122}$Sn and the  
incident proton energy is equal 20 MeV. Right: Comparison between the calculated and 
the experimental cross sections for the transition to the first excited 0$^+$ state of the final
 nucleus. The 
reaction is $^{124}$Sn($p,t$)$^{122}$Sn and the  
incident proton energy is equal 35 MeV. }
\label{fig3}
\end{figure}

\begin{figure}[htb]
\begin{center}
\parbox[c]{16cm}{\includegraphics[width=14cm]{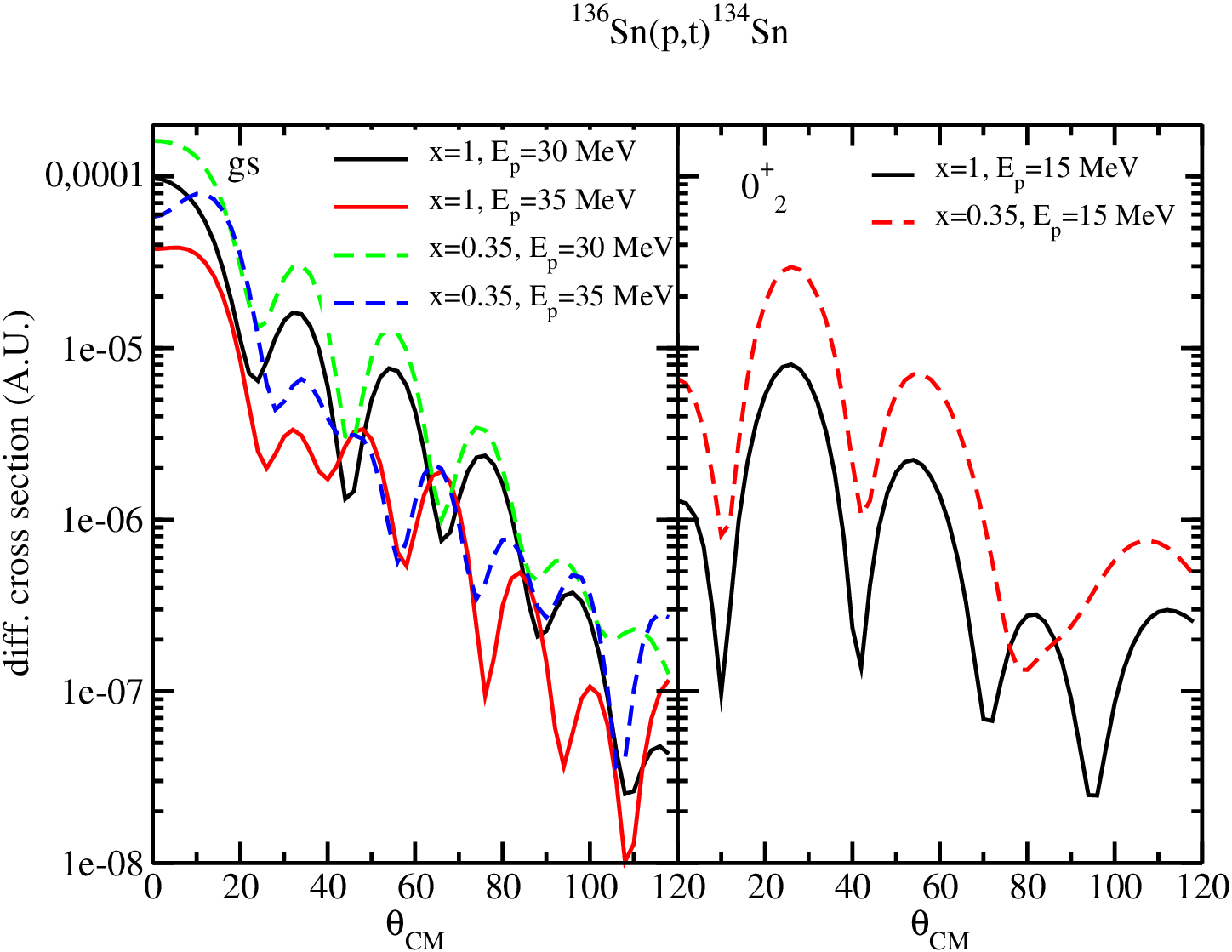}}
\end{center}
\caption{\small (Color online) Left: Cross sections for the reaction 
$^{136}$Sn($p,t$)$^{134}$Sn obtained in the cases of a pure surface 
interaction ($x=$ 1) and a mixed interaction ($x=$ 0.35). The transition to 
the ground state of the final nucleus is considered and two values of the incident proton 
energy are selected, $E_p =$ 30 and 35 MeV. Right: Cross sections for the reaction 
$^{136}$Sn($p,t$)$^{134}$Sn obtained in the cases of a pure surface 
interaction ($x=$ 1) and a mixed interaction ($x=$ 0.35). The transition to 
the first excited 0$^+$ state of the final nucleus is considered and the value of 15 MeV is selected for the incident proton 
energy.  }
\label{fig4}
\end{figure}

\begin{figure}[htb]
\begin{center}
\parbox[c]{16cm}{\includegraphics[width=14cm]{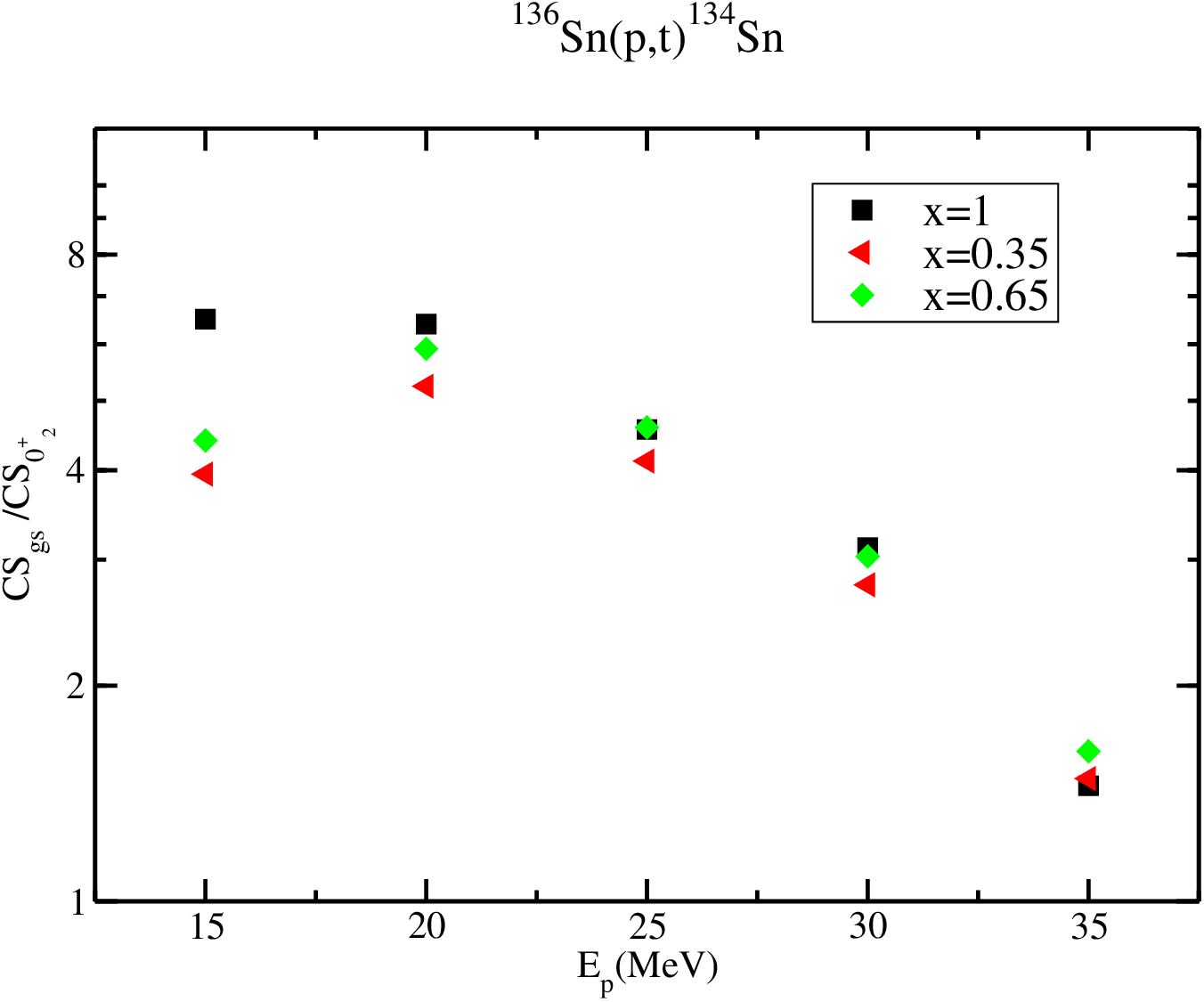}}
\end{center}
\caption{\small (Color online)   Ratios of the cross sections associated to the 
$gs$ and to the 0$^+_2$ transitions at different proton energies 
for the reaction $^{136}$Sn($p,t$)$^{134}$Sn. }
\label{fig5}
\end{figure}

\end{document}